\documentclass[preprint,aps]{revtex4-2}
\usepackage{mathrsfs}
\usepackage{graphicx}
\usepackage{multirow}
\usepackage{makecell}
\usepackage{booktabs}
\usepackage{color}
\usepackage{bm}

\usepackage{hyperref}
\hypersetup{
    colorlinks,
    citecolor=black,
    filecolor=black,
    linkcolor=black,
    urlcolor=black
}
\begin{document}

\title{Non-Hermitian Boundary in a Surface Selective Reconstructed Magnetic Weyl Semimetal}

\author{Cong Li$^{1,\sharp,*}$, Yang Wang$^{1,\sharp}$, Jianfeng Zhang$^{2,\sharp}$, Hongxiong Liu$^{2,\sharp}$, Wanyu Chen$^{1}$, Guowei Liu$^{3}$, Hanbin Deng$^{3}$, Craig Polley$^{4}$, Balasubramanian Thiagarajan$^{4}$, Timur Kim$^{5}$, Jiaxin Yin$^{3}$, Youguo Shi$^{2}$, Tao Xiang$^{2,*}$, Oscar Tjernberg$^{1,*}$
}

\affiliation{
\\$^{1}$Department of Applied Physics, KTH Royal Institute of Technology, Stockholm 11419, Sweden
\\$^{2}$Beijing National Laboratory for Condensed Matter Physics, Institute of Physics, Chinese Academy of Sciences, Beijing 100190, China
\\$^{3}$Department of Physics, Southern University of Science and Technology, Shenzhen, Guangdong 518055, China
\\$^{4}$MAX IV Laboratory, Lund University, 22100 Lund, Sweden
\\$^{5}$Diamond Light Source, Harwell Campus, Didcot, OX11 0DE, United Kingdom
\\$^{\sharp}$These people contributed equally to the present work.
\\$^{*}$Corresponding authors: conli@kth.se, txiang@iphy.ac.cn, oscar@kth.se
}

\pacs{}

\maketitle


\begin{center}
{\bf Abstract}
\end{center}

{\bf Non-Hermitian physics, studying systems described by non-Hermitian Hamiltonians \cite{NHatano_PRL1996_DRNelson,CMBender_RPP2007,REGanainy_NP2018_DNChristodoulides,YAshida_AIP2021_MUeda,EJBergholtz_RMP2021_FKKunst,KDing_NRP2022_GMa,RLin_FP2023_CHLee,SYYao_PRL2018_ZWang,NOkuma_AR2023_MSato}, reveals unique phenomena not present in Hermitian systems\cite{NHatano_PRL1996_DRNelson,CMBender_RPP2007,REGanainy_NP2018_DNChristodoulides,YAshida_AIP2021_MUeda,EJBergholtz_RMP2021_FKKunst,KDing_NRP2022_GMa,RLin_FP2023_CHLee,SYYao_PRL2018_ZWang,NOkuma_AR2023_MSato,YXiong_JPC2018,ZSYang_PRL2020_1_JPHu,ZSYang_PRL2020_2_JPHu,YFYi_PRL2020_ZSYang,LHLi_NC2020_JBGong,NOkuma_PRL2020_MSato,XJZhang_AIPX2022_YFChen}. Unlike Hermitian systems, non-Hermitian systems have complex eigenvalues, making their effects less directly observable. Recently, significant efforts have been devoted to incorporating the non-Hermitian effects into condensed matter physics\cite{TYoshida_PRB2018_NKawakami,AAZyuzin_PRB2018_AYZyuzin,TMPhilip_PRB2018_MJGilbert,MPapaj_PRB2019_LFu,TYoshida_PRB2019_YHatsugai,KMoors_PRB2019_TLSchmidt,AAZyuzin_PRB2019_PSimon,EJBergholtz_PRR2019_JCBudich}. However, progress has been hindered by the absence of a viable experimental approach. Here, the discovery of the surface-selectively spontaneous reconstructed Weyl semimetal NdAlSi provides a feasible experimental platform for studying non-Hermitian physics. Utilizing angle-resolved photoemission spectroscopy measurements, surface-projected density functional theory calculations, and scanning tunneling microscopy measurements, we demonstrate that surface reconstruction in NdAlSi alters surface Fermi arc connectivity and generates new isolated non-topological surface Fermi arcs by introducing non-Hermitian terms. The surface-selective spontaneous reconstructed Weyl semimetal NdAlSi can be viewed as a Hermitian bulk -- non-Hermitian boundary system. The isolated non-topological surface Fermi arcs on the reconstructed surface act as a loss mechanism and open boundary condition for the topological electrons and bulk states, serving as non-Hermitian boundary states. This discovery provides a good experimental platform for exploring new physical phenomena and potential applications based on boundary non-Hermitian effects, extending beyond purely mathematical concepts.
}


Quantum mechanics is well established for studying isolated quantum systems under one of the fundamental assumptions that all observable operators are Hermitian. However, for the study of an open system with non-conservative or non-equilibrium, non-Hermitian theory provides an excellent framework relative to Hermitian theories\cite{NHatano_PRL1996_DRNelson,CMBender_RPP2007,REGanainy_NP2018_DNChristodoulides,YAshida_AIP2021_MUeda,EJBergholtz_RMP2021_FKKunst,KDing_NRP2022_GMa,RLin_FP2023_CHLee,SYYao_PRL2018_ZWang,NOkuma_AR2023_MSato,YXiong_JPC2018,ZSYang_PRL2020_1_JPHu,ZSYang_PRL2020_2_JPHu,YFYi_PRL2020_ZSYang,LHLi_NC2020_JBGong,NOkuma_PRL2020_MSato,XJZhang_AIPX2022_YFChen}. In a non-Hermitian system, the interplay between gain and loss can give rise to novel phenomena that are fundamentally different from those in Hermitian systems, such as exceptional points (EPs) and the non-Hermitian skin effect (NHSE)\cite{SYYao_PRL2018_ZWang,NOkuma_AR2023_MSato,LHLi_NC2020_JBGong,NOkuma_PRL2020_MSato,XJZhang_AIPX2022_YFChen}. In recent years, the study of EPs and the associated non-Hermitian dynamics has garnered significant attention, especially in the areas of optics systems\cite{SBLee_PRL2009_JHLee,AGuo_PRL2009_DNChristodoulides,CERuter_NP2010_DKip,MLiertzer_PRL2012_SRotter}, photonic crystals\cite{ARegensburger_Nature2012_UPeschel,BZhen_Nature2015_MSoljacic,HYZhou_Science2018_BZhen,MAMiri_Science2019_AAlu} and electrical circuits\cite{THelbig_NP2020_RThomale,DYZou_NC2021_XDZhang,HYuan_AS2023_XDZhang}. Currently, considerable attention is focused on incorporating the non-Hermitian effects into condensed matter physics\cite{TYoshida_PRB2018_NKawakami,AAZyuzin_PRB2018_AYZyuzin,TMPhilip_PRB2018_MJGilbert,MPapaj_PRB2019_LFu,TYoshida_PRB2019_YHatsugai,KMoors_PRB2019_TLSchmidt,AAZyuzin_PRB2019_PSimon,EJBergholtz_PRR2019_JCBudich}, though experimental breakthroughs have yet to be achieved. Here, we find that the surface reconstructed Weyl semimetal can serve as a viable experimental platform for observing non-Hermitian effects. In Weyl semimetals, surface reconstruction will disrupt the translational symmetry of the crystal structure at the boundary between the bulk and surface, creating exotic physical properties, including a non-Hermitain effect at the boundary. However, so far due to the lack of suitable material systems\cite{ABedoya-Pinto_AM2021_SSPParkin,QZou_NPJ2022_ZGai}, the properties of surface reconstructed Weyl semimetals are rarely discussed.

Using angle-resolved photoemission spectroscopy (ARPES), scanning tunneling microscopy (STM) measurements and density functional theory (DFT) calculations on the noncentrosymmetric magnetic Weyl semimetal NdAlSi\cite{CLi_NC2023_OTjernberg,CLi_arxiv2024_OTjernberg,JGaudet_NM2021_CLBroholm}, we demonstrate that spontaneous reconstruction occurs on the Al-terminated surface after sample cleaving, but not on the complementary Nd-terminated surface. This provides an opportunity to explore unique physical phenomena in surface reconstructed Weyl semimetals in contrast to unreconstructed surfaces. Such phenomena include topological phase transitions, symmetry-breaking effects, and the interplay between surface and bulk topological properties, as well as the interaction between magnetism and surface reconstruction in Weyl semimetals. Compared with conventional Weyl semimetals, the surface reconstructed Weyl semimetals display a broader range of physical phenomena. The reconstructed surface leads to a modified surface Hamiltonian, impacting the topology of surface states by changing the dispersion and connectivity of surface Fermi arcs (SFAs), shifting their positions, or introducing new topological features. The surface reconstruction can also influence the coupling between surface and bulk states, potentially resulting in surface and bulk hybridized states with unique transport properties and new types of surface-bound excitations. The occurrence of a surface reconstruction can even affect the bulk-boundary correspondence, an important principle in condensed matter physics\cite{GMGraf_CMP2013_MPorta,YKubota_CMP2017,KHashimoto_PTEP2017_XWu,KGomi_PTEP2022}. Furthermore, surface reconstruction alters the boundary conditions at the surface of the Weyl semimetals. This modification can further introduce non-Hermitian terms into the effective Hamiltonian that describes the surface states, giving rise to non-Hermitian effects at the boundary\cite{NHatano_PRL1996_DRNelson,CMBender_RPP2007,REGanainy_NP2018_DNChristodoulides,YAshida_AIP2021_MUeda,EJBergholtz_RMP2021_FKKunst,KDing_NRP2022_GMa,RLin_FP2023_CHLee,SYYao_PRL2018_ZWang,NOkuma_AR2023_MSato,YXiong_JPC2018,ZSYang_PRL2020_1_JPHu,ZSYang_PRL2020_2_JPHu,YFYi_PRL2020_ZSYang,LHLi_NC2020_JBGong,NOkuma_PRL2020_MSato,XJZhang_AIPX2022_YFChen}. These non-Hermitian effects will lead to novel physical properties and behaviors not present in purely Hermitian systems, opening new avenues for research and applications in topological materials.

In this work, we present ARPES measurements and DFT calculations to systematically investigate the electronic structure and topological properties of Weyl semimetal NdAlSi, with a focus on how these properties are influenced by surface reconstruction. Our findings reveal that surface reconstruction alters the connectivity of surface Fermi arcs (SFAs) in NdAlSi by introducing non-Hermitian terms into the effective Hamiltonian for surface states, and leading to the emergence of new isolated non-topological SFAs (NTSFAs) that do not connect to chirally opposite Weyl points and further hybridize with bulk states. The non-Hermitian effects associated with the reconstructed surface also causes the SFAs to broaden compared to the unreconstructed surface. In the presence of a magnetic field, the NTSFAs act as a loss mechanism and open boundary condition (OBC) for topological electrons and bulk states, resulting in a boundary non-Hermitian effect. This discovery demonstrates that the surface-selectively spontaneous reconstructed Weyl semimetal NdAlSi can be viewed as a Hermitian bulk -- non-Hermitian boundary system, offering an exceptional platform for exploring new physical phenomena and developing innovative technologies based on boundary non-Hermitian effects.

\begin{figure*}[tbp]
\begin{center}
\includegraphics[width=1\columnwidth,angle=0]{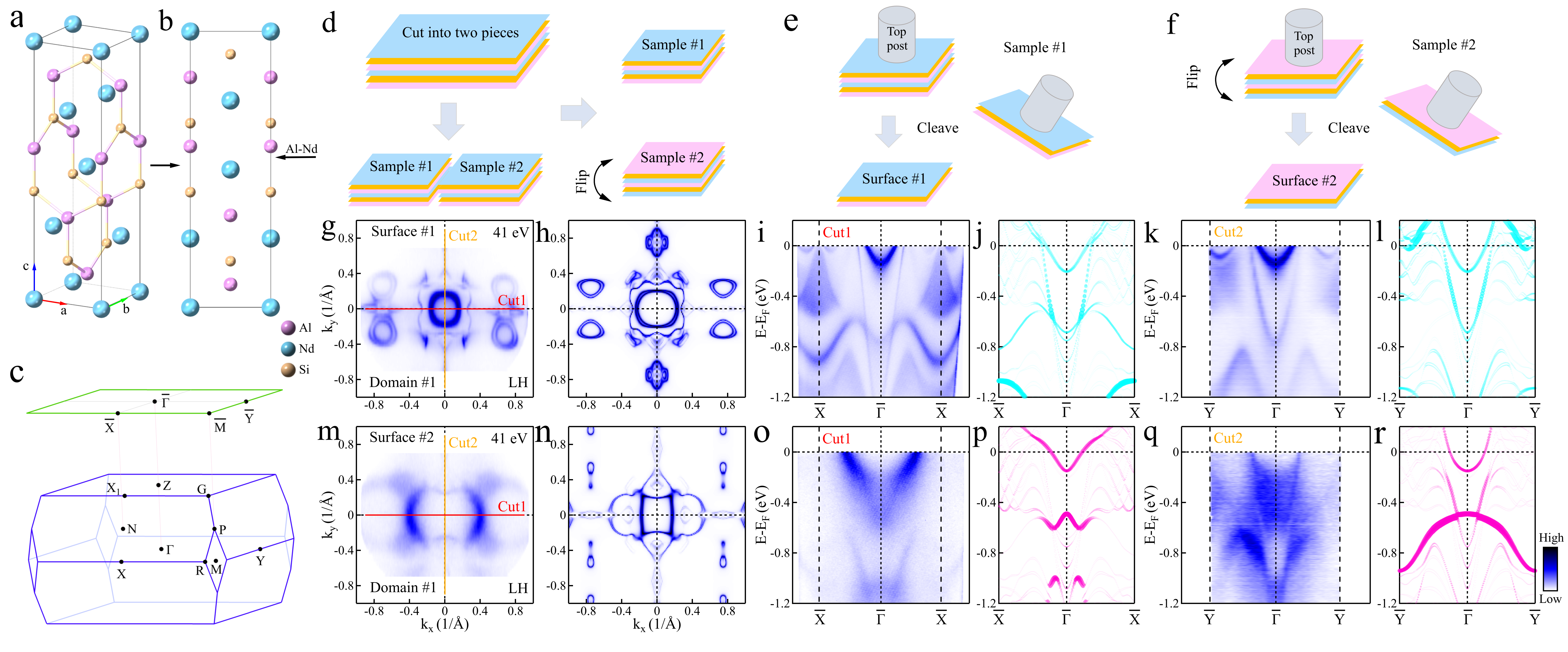}
\end{center}
\caption{\footnotesize\textbf{Determination of the surface termination in NdAlSi.} (a) The crystal structure of NdAlSi with the space group $I4_{1}md$ (no. 109). (b) The side view of (a). (c) The 3D Brillouin zone (BZ) of the original unit cell of NdAlSi, and the corresponding two-dimensional BZ projected on the (001) plane (green lines) in the pristine phase in (a). (d-f) The schematic diagram for get the different termination surfaces of NdAlSi. (g) Fermi surface of NdAlSi measured on surface 1 with photon energy of 41 eV under LH polarization. (h) Surface projected DFT calculations of the (001) Fermi surface on the Nd atom terminated surface cleavage at the Al-Nd layer. (i) and (k) The band dispersions along the path of Cut1 (i) and Cut2 (k) in (g) measured on surface 1 with photon energy of 41 eV under LH polarization. (j) and (l) The corresponding surface projected DFT calculations of the bands along $\overline{X}-\overline{\Gamma}-\overline{X}$ (j) and $\overline{Y}-\overline{\Gamma}-\overline{Y}$ (l) directions of the terminal surface of Nd atom cleavage at the Al-Nd layer. (m,o,q) The similar measurements as (g,i,k) but from surface 2. (n) Surface projected DFT calculations of the (001) Fermi surface on the Al atom terminated surface cleavage at the Al-Nd layer. (p) and (r) The corresponding surface projected DFT calculations of the bands along $\overline{X}-\overline{\Gamma}-\overline{X}$ (p) and $\overline{Y}-\overline{\Gamma}-\overline{Y}$ (r) directions of the terminal surface of Al atom cleavage at the Al-Nd layer. There are two mutually orthogonal domain structures on the cleavage surface of NdAlSi. The above measurements come from domain 1.
}
\label{1}
\end{figure*}

Firstly, we demostrate the existence of a surface-selective spontaneous reconstruction in the Weyl semimetal NdAlSi. The results of multiple measurements (see section 1 in the Supplementary Material for details) have revealed the existence of two types of Fermi surface structures in NdAlSi, as shown in Fig.~\ref{1}g and~\ref{1}m. Since NdAlSi is a noncentrosymmetric crystal, its upper and lower surfaces are asymmetrical. Therefore, it is reasonable to assume that these two Fermi surfaces may originate from different termination surfaces (the upper and lower surfaces) of NdAlSi after cleavage. To confirm this conjecture, measurements were performed on these two distinct terminal surfaces. Firstly, a large sample was cut into two pieces, labeled sample 1 and sample 2, respectively (Fig.~\ref{1}d). Sample 2 was flipped 180 degrees relative to sample 1. Both samples were then mounted with top posts as shown in Fig.~\ref{1}e-\ref{1}f and cleaved to obtain surface 1 (Fig.~\ref{1}e) and surface 2 (Fig.~\ref{1}f). Fig.~\ref{1}g shows the Fermi surface map of surface 1, measured with a photon energy of 41 eV under linear horizontal (LH) polarization (the polarization dependent measurements see section 2 in the Supplementary Material). This Fermi surface shape has been previously observed in the RAlX (R: Rare earth; X: Si or Ge) family\cite{DSSanchez_NC2020_MZHasan,RLuo_PRB2023_SBorisenko,APSakhya_PRM2023_MNeupane,YCZhang_CP2023_MYi,CLi_NC2023_OTjernberg,CLi_arxiv2024_OTjernberg}. The well agreement between surface-projected DFT calculations of the Nd-terminated surface cleavage at the Al-Nd layer (Fig.~\ref{1}h,~\ref{1}j and~\ref{1}l) and measurements of surface 1 (Fig.~\ref{1}g,~\ref{1}i and~\ref{1}k), indicating the Nd terminal surface nature of surface 1\cite{CLi_NC2023_OTjernberg}. When we make measurements on surface 2, the measured Fermi surface is indeed the other type of Fermi surface. Therefore, the measured Fermi surface in Fig.~\ref{1}m should be derived from the terminal surface of Al atom cleavage at the Al-Nd layer. However, the measurements of surface 2 (Fig.~\ref{1}m,~\ref{1}o and~\ref{1}q) does not align well with the surface-projected DFT calculations of Al atom cleavage at the Al-Nd layer (Fig.~\ref{1}n,~\ref{1}p and~\ref{1}r) (the comparison of constant energy contours see section 3 in the Supplementary Material).

\begin{figure*}[tbp]
\begin{center}
\includegraphics[width=1\columnwidth,angle=0]{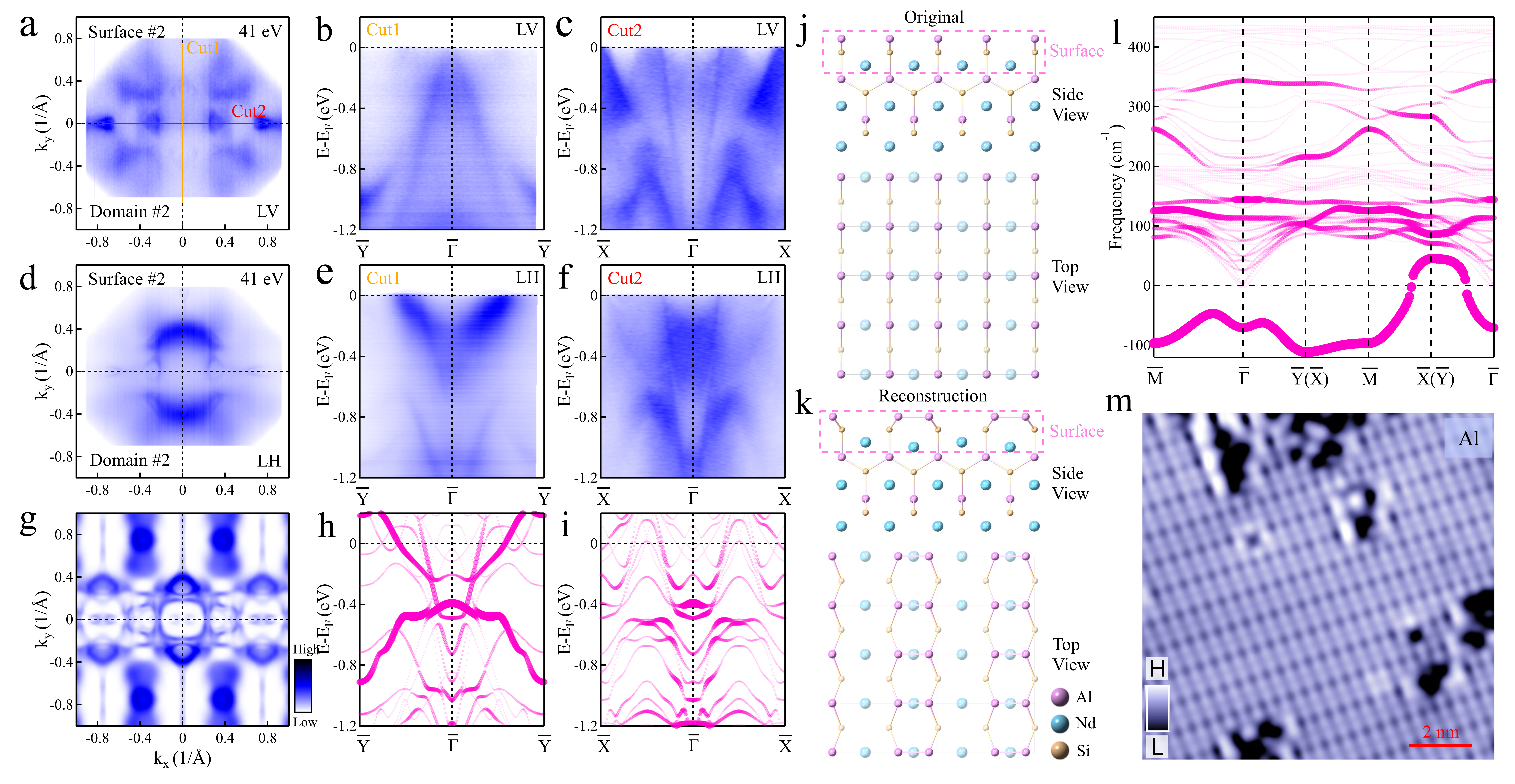}
\end{center}
\caption{\footnotesize\textbf{Surface state reconstruction of the Al atom terminated surface cleavage at the Al-Nd layer.} (a) Fermi surface of NdAlSi measured on surface 2 with photon energy of 41 eV under LV polarization. (b-c) The band dispersions along the path of Cut1 (b) and Cut2 (c) in (a) measured on surface 2 with photon energy of 41 eV under LV polarization. (d-f) The similar measurements as (a-c) but under LH polarization. (g) Surface projected DFT calculations of the (001) Fermi surface on the Al atom terminated surface cleavage at the Al-Nd layer with consideration of surface reconstruction. (h-i) The corresponding surface projected DFT calculations of the bands along $\overline{Y}-\overline{\Gamma}-\overline{Y}$ (h) and $\overline{X}-\overline{\Gamma}-\overline{X}$ (i) directions of the terminal surface of Al atom cleavage at the Al-Nd layer with consideration of surface reconstruction. The surface projected DFT calculated band dispersion along high-symmetry directions across the BZ of the terminal surface of Al atom cleavage at the Al-Nd layer with consideration of surface reconstruction see Fig. S11 in the Supplementary Material for detailed. (j-k) Crystal structure of NdAlSi before (j) and after (k) surface reconstruction. (l) The surface projected phonon spectrum calculations on the Al terminated surfaces at the Al-Nd layer. The spectral intensity is denoted by the size and transparency of the markers. The higher the intensity, the greater the proportion of surface projection. (m) STM images of NdAlSi measured on the terminal surface of Al atom cleavage at the Al-Nd layer at 4.7 K. Scan conditions: 0.5 V, 0.1 nA. Negative bias implies filled states.
}
\label{2}
\end{figure*}

To understand the reasons behind this difference, further data was recorded. The results reveal two mutually orthogonal domain structures on the cleaved surface of NdAlSi (see section 4 in the Supplementary Material for details)\cite{CLi_NC2023_OTjernberg,CLi_arxiv2024_OTjernberg}. Fig.~\ref{2}a-\ref{2}f show the polarization dependent measurements of domain 2 measured on surface 2 with a photon energy of 41 eV. Hole-like bands are observed near the Fermi level in the BZ center, as shown in Fig.~\ref{2}b. However, surface-projected DFT band calculations do not show any hole-like bands near the Fermi level at the BZ center, regardless of surface termination (see section 5 in the Supplementary Material for details). But it appears near the $\rm\overline{X}$ and $\rm\overline{Y}$ points in the BZ of NdAlSi, suggesting a possible surface reconstruction on surface 2 of NdAlSi. Then, we performed surface projected phonon spectrum calculations on the Nd (see section 6 in Supplementary Material for details) and Al (Fig.~\ref{2}l) terminated surfaces respectively. A negative frequency on the Al-terminated surface (surface 2, Fig.~\ref{2}l) but not on the Nd-terminated surface (surface 1), indicating that the Al-terminated surface (surface 2, Fig.~\ref{2}j) is unstable and prone to surface reconstruction\cite{JFZhang_PRB2021_ZYLu}. Further calculations indicate that the surface energy of surface 2 is the lowest when considering a 2$\times$1 surface reconstruction (Fig.~\ref{2}k). Surface projected DFT calculations (Fig.~\ref{2}g-\ref{2}i) on the Al-terminated surface were then repeated, assuming a 2$\times$1 surface reconstruction (Fig.~\ref{2}k). By comparison with the measurements on surface 2 of NdAlSi (Fig.~\ref{2}a-\ref{2}f), it is found that both the Fermi surface and the band structure are well matched. The comparisons of the constant energy contours at various binding energies see section 7 in the Supplementary Material. In addition, the atomic morphologies of surface 1 (see section 8 in Supplementary Material for details) and surface 2 were determined by STM as shown in Fig.~\ref{2}m. The STM measurements also reveal a 2$\times$1 surface reconstruction on surface 2 but not on surface 1, confirming the ARPES measurements and the surface-projected DFT calculations. 

\begin{figure*}[tbp]
\begin{center}
\includegraphics[width=0.7\columnwidth,angle=0]{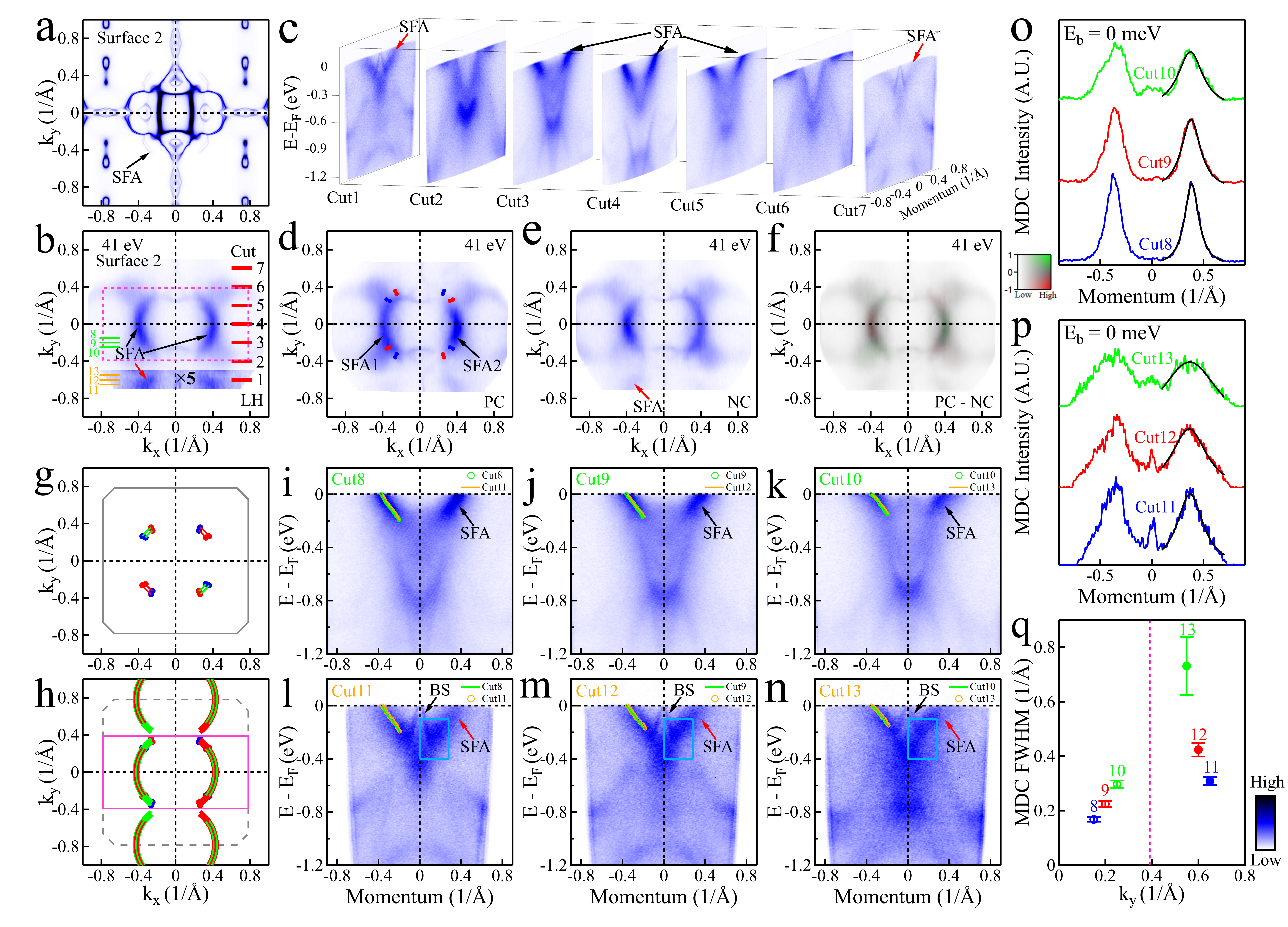}
\end{center}
\caption{\footnotesize\textbf{SFA reconstruction on the Al atom terminated surface cleavage at the Al-Nd layer.} For ease of comparison and description, Fig.~\ref{1}n and Fig.~\ref{1}m are replotted in (a) and (b). (c) Band structures of NdAlSi measured on surface 2 with photon energy of 41 eV under LH polarization. Cut 1-7 correspond to the energy-momentum sections of (b) at k$_y$ = -0.6, -0.4, -0.2, 0, 0.2, 0.4, 0.6 $\AA^{-1}$, respectively. (d-e) Fermi surface of NdAlSi measured on surface 2 with photon energy of 41 eV under PC and NC polarizations. (f) The CD Fermi surface obtained by the difference between (d) and (e). The spectrum is plotted with the two-dimensional colorscale shown, with the horizontal axis corresponding to intensity and the vertical axis corresponding to CD signal. (g-h) SFA schematic on surface 2 without (g) and with (h) considering the surface reconstruction. Gray lines in (g) and pink lines in (h) are the first BZ boundary at the k$_z$ = 0 $\pi/c$ plane without and with considering the surface reconstruction. The red dots representing nodes with chiralities +1 and blue dots representing -1. The SFAs marked in red and green correspond to different connectivity of SFAs. (i-n) Band dispersions measured along Cut8-Cut13 in (b) with photon energy of 41 eV under LH polarization. The green (orange) hollow points in (i-k) [(l-n)] mark the band dispersions obtained by fitting energy dependent MDCs of corresponding band structures. For ease of comparison, the fitted band dispersions of Cut11-Cut13 are replotted in Cut8-Cut10 with orange sloid lines, the fitted band dispersions of Cut8-Cut10 are replotted in Cut11-Cut13 with green sloid lines. The area surrounded by the cyan lines represent the hybrid area of SFAs and bulk states (BS). (o) and (p) The extracted MDCs from (i-k) and (l-n) at Fermi level. (q) The FWHM of MDCs in (o-p) obtained by fitting these MDCs with Lorentz curves (black lines). The scattering rate of quasiparticle can be obtained by multiplying the FWHM of MDCs by the corresponding Fermi velocities.
}
\label{3}
\end{figure*}

Next, we will focus on the influence of surface reconstruction on topological properties. The surface reconstruction on surface 2 causes the corresponding Fermi surface (Fig.~\ref{3}b) to undergo reconstruction as well. In the reconstrucetd Fermi surface, two arc features with strong spectral weight (marked by black arrows) are clearly observed. The photon energy dependence of the arc features is detailed in section 10 of the Supplementary Material, showing no significant dispersion along the k$_z$ direction, indicating the surface state nature of them. To visualize the band dispersion of the arc features, band structure cuts across the arc features were measured as shown in Fig.~\ref{3}c. Fig.~\ref{3}d and~\ref{3}e show similar measurements to Fig.~\ref{3}b but under positive circular (PC, Fig.~\ref{3}d) and negative circular (NC, Fig.~\ref{3}e) polarizations. The polarization dependence indicates that the arc feature observed in Fig.~\ref{3}b is a mixture of two different lengths of arcs. The confirmation of the arc features in Fig.~\ref{3}b are the SFAs see section 11 in Supplementary Material. The circular-dichroic (CD) measurements are shown in Fig.~\ref{3}f, revealing that the longer SFAs are very close to the the shorter SFAs (Fig.~\ref{3}h). For detailed connectivity of SFAs see section 12 in Supplementary Material. It is clear that the surface reconstruction alters the connectivity of SFAs (Fig.~\ref{3}g-\ref{3}h). Notably, while surface reconstructions have been observed in the Weyl semimetals NbP\cite{ABedoya-Pinto_AM2021_SSPParkin} and BaMnSb$_2$\cite{QZou_NPJ2022_ZGai}, no SFA reconstruction is observed in these materials. As a matter of fact, no SFAs are observed in BaMnSb$_2$\cite{HTRong_CPB2021_XJZhou}.

Due to the 2$\times$1 surface reconstruction of surface 2, the surface cell size doubles along the reconstruction direction, causing the corresponding BZ to fold to half its original size in that direction, as depicted in Fig.~\ref{3}h. This folding creates new SFAs to satisfy the new translational symmetry of the surface. This phenomenon is directly observed in the ARPES measurements as shown in Fig.~\ref{3}b-\ref{3}f and Fig.~\ref{3}i-\ref{3}n. However, since the reconstruction occurs only on surface 2, the surface states of surface 1 and the bulk states remain unchanged\cite{CLi_NC2023_OTjernberg,CLi_arxiv2024_OTjernberg} (see section 13 in Supplementary Material). The inconsistency of surface 2 with the lattice periodicity of the bulk structure, breaks both in-plane and out-of-plane translational symmetries when electrons transition from the bulk to the surface. This disruption results in the decoupling of surface and bulk states on surface 2. Consequently, the new SFAs generated from the surface reconstruction on surface 2 cannot connect to the pair of Weyl points with opposite chirality, indicating that they are isolated NTSFAs, as illustrated in Fig.~\ref{3}h. However, due to the translational symmetry of the surface, the NTSFAs exhibit the same dispersion as the original SFAs, as shown in Fig.~\ref{3}i-\ref{3}n. Additionally, these NTSFAs hybridize with the bulk states (Fig.~\ref{3}l-\ref{3}n), altering the electronic structure of the bulk states. Furthermore, the full width at half maximum (FWHM) of the momentum distribution curves (MDCs) at the Fermi level (Fig.~\ref{3}o-\ref{3}q) shows that the closer the end point of the SFA is, the shorter its lifetime. Moreover, the lifetime of the NTSFAs is shorter than that of the original SFAs (Fig.~\ref{3}q), indicating higher dissipation.

\begin{figure*}[tbp]
\begin{center}
\includegraphics[width=1\columnwidth,angle=0]{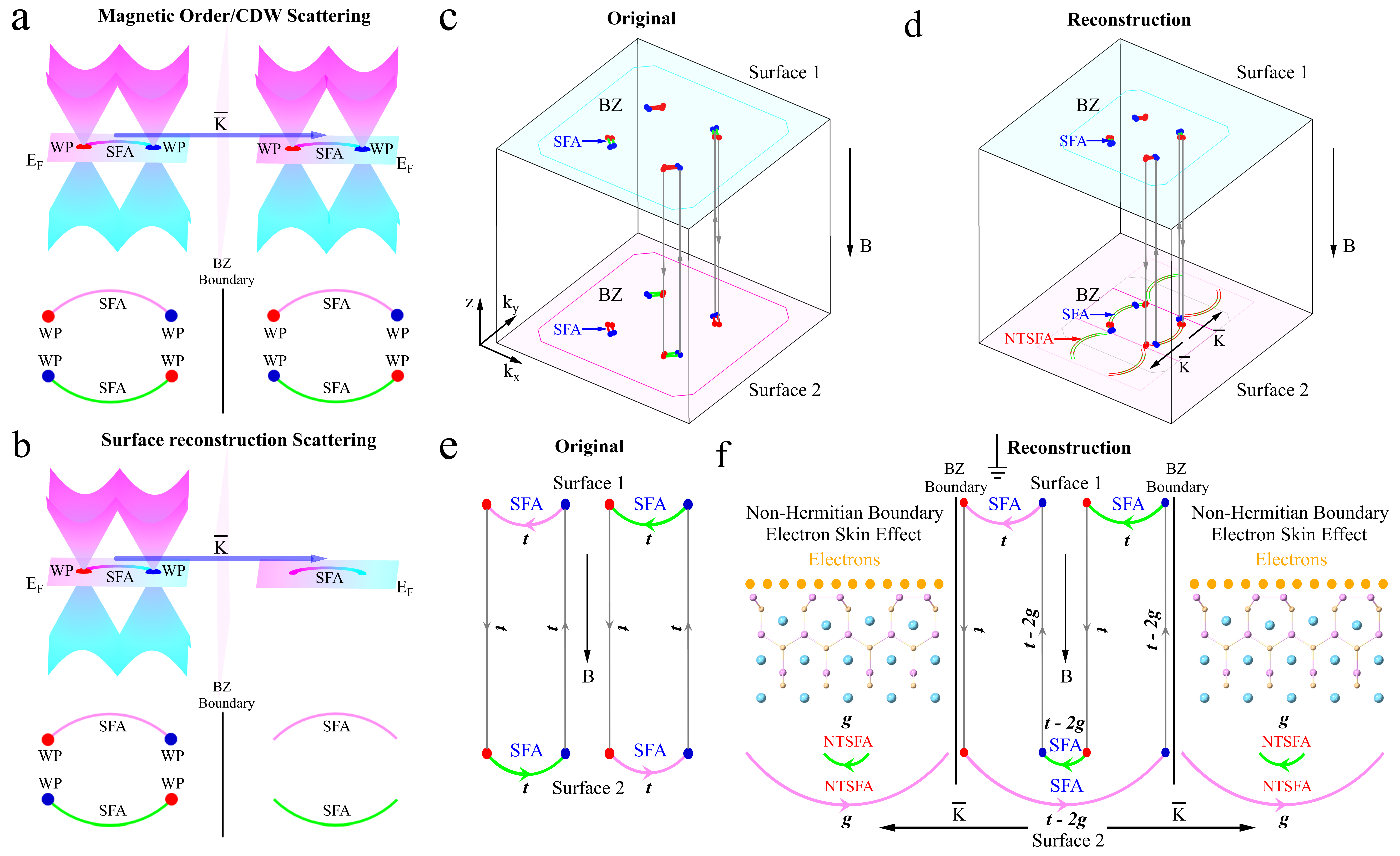}
\end{center}
\caption{\footnotesize\textbf{Schematic diagram of the new SFAs resulting from surface reconstruction.} (a) Schematic diagram of the original Weyl points (WPs) and SFAs being scattered to the new location of BZ due to the folding of the 3D BZ. (b) Schematic diagram of the original SFAs being scattered to the new location of BZ due to the folding of the surface BZ induced by surface reconstruction. (c-d) Schematic illustration of Weyl orbits that weave together chiral bulk states and SFA of top (surface 1) and bottom (surface 2) surfaces without (c) and with (d) considering the surface reconstruction on surface 2, while the SFAs of surface 1 are well studied in Ref. \cite{CLi_NC2023_OTjernberg}. We have simplified them here for the sake of description. The above Weyl nodes are the DFT calculated results\cite{CLi_NC2023_OTjernberg}. (e-f) Electrons exhibit an unusual path in real and momentum ($z$-$k_x$-$k_y$) space under an external magnetic field along the $z$ direction when considering (e) and without considering (f) the surface reconstruction on surface 2. The gray arrows show the real-space motion of the electrons between the top (surface 1) and the bottom (surface 2) surfaces. The green and pink arrows show the electron’s momentum space trajectories tracing out the constant energy contour of the Fermi arcs on the surfaces.  
}
\label{4}
\end{figure*}

In general, when BZ folding is induced by charge density waves (CDW) or magnetic order, bulk linear bands, Weyl points, and topological SFAs can be relocated to new positions in the BZ, as depicted in Fig.~\ref{4}a. However, when BZ folding results from surface reconstruction, only the topological SFAs are relocated to new positions in the BZ, leading to the emergence of NTSFAs as shown in Fig.~\ref{4}b. According to the bulk-boundary correspondence\cite{GMGraf_CMP2013_MPorta,YKubota_CMP2017,KHashimoto_PTEP2017_XWu,KGomi_PTEP2022}, the SFA and bulk state intersect only at the Weyl points, implying that in transport measurements, electrons on the surface can only communicate with bulk electrons through Weyl points. However, due to the presence of NTSFA and bulk states hybridization in the surface reconstructed Weyl semimetal (Fig.~\ref{3}l-\ref{3}n), the surface electrons have more pathways to interact with bulk electrons, suggesting the possibility of novel transport phenomena. The NTSFA observed here is a new kind of non-topological Fermi arc, completely different from the Fermi arc observed in cuprate underdoped superconductors, which is associated with the pseudogap in the normal state\cite{HDing_Nature1996_JGiapintzakis,AGLoeser_Science1996_AKapitulnik,MRNorman_Nature1998_DGHinks}, and also distinct from the recently discovered Fermi arc resulting from antiferromagnetic splitting\cite{BSchrunk_Nature2022_AKaminski}.

Interestingly, the discovery of surface-selective reconstruction in the magnetic Weyl semimetal NdAlSi provides a fertile ground for studying non-Hermitian topological physics. Surface reconstruction introduces new periodicities and surface potentials, which couple to the surface Hamiltonian with non-Hermitian terms, modifying the boundary conditions. The new surface potential alters SFA connectivity and also leads to asymmetric electron hopping ($\delta H$) in momentum space, causing the emergence of NTSFAs, as shown in Fig.~\ref{3}l-\ref{3}n. If the electron hopping is symmetric, the spectral weight of the NTSFA is zero. Due to the complex surface potentials induced by surface reconstruction [$i\Gamma(x)$, reconstruct in the $x$ direction], which will modify the surface Green function or surface state eigenvalue equation to include complex eigenvalues [$E(k) = E_{Re}(k) + iE_{Im}(k)$] from the non-Hermitian terms. In this case, the Hamiltonian of non-Hermitian terms introduced by the surface reconstruction can be written as $H_{NH} = i\Gamma(x) + \delta H$. The imaginary part of the eigenvalues shorthen the lifetime of the SFA states. The change in Fermi arc quasiparticle lifetime due to the non-Hermitian terms differs from that caused by impurity scattering. For non-Hermitian effects, the closer to the Weyl point, the shorter the quasiparticle lifetime, as the interaction between the new surface potential and topological electrons strengthens near the Weyl point. In contrast, for impurity scattering, topological protection near the Weyl point makes quasiparticles more robust to random scattering, thus shorter lifetimes for Fermi arc quasiparticles farther from the Weyl point. The non-Hermitian effect induces a global change in the surface, whereas impurity scattering leads to only local changes.

In NdAlSi, the quasiparticle lifetime of SFA states on the unreconstructed surface is at least 4 to 11 times longer than on the reconstructed surface (see section 14 in the Supplementary Material for details). According to the results in Fig.~\ref{3}q, it is apparent that the increased quasiparticle decay or dissipation in the SFA on the reconstructed surface is linked to the non-Hermitian term introduced by the surface reconstruction, affecting all points across the reconstructed surface. The electron scattering between the NTSFA and SFA further shortens the electron lifetime in the NTSFA, as shown in Fig.~\ref{3}o-\ref{3}q. Additionally, the NTSFAs due to asymmetric electron hopping can act as a loss mechanism for the bulk states, which is essential for introducing non-Hermitian effects. When an external magnetic field is applied, if there is no surface reconstruction in the NdAlSi, a closed loop current forms by the topological electrons in the real and momentum ($z$-$k_x$-$k_y$) space as shown in Fig.~\ref{4}c and~\ref{4}e. However, due to the presence of surface reconstruction on the Al-terminated surface, the trajectories of electrons in real and momentum space change, generating the NTSFA simultaneously, as shown in Fig.~\ref{4}d and~\ref{4}f. When topological electrons flow through surface 2, some will return to the bulk along the new path, while others will be scattered to the NTSFA, becoming non-topological electrons. Topological electrons returning to the bulk through a new path with a shorter lifetime will increase the scattering of bulk state electrons, degrading the topological transport properties. Moreover, the scattered electrons on the NTSFA have even shorter lifetimes (Fig.~\ref{3}p-\ref{3}q) and are unable to form closed loops under a magnetic field (Fig.~\ref{4}f). As a result, the dissipated electrons eventually accumulate on the reconstructed surface, establishing a new equilibrium between surface electron accumulation and dissipation.

Throughout the entire process, the NTSFAs due to asymmetric electron hopping in momentum space serve as a loss mechanism for the topological electrons, resulting in the number of topologically protected electrons no longer being conserved. Additionally, for the bulk electrons, the presence of the NTSFAs functions as an OBC, effectively funneling bulk electrons to the reconstructed surface, where they accumulate. In this scenario, the NdAlSi single crystal acts as a generator, producing a net electron current (2$g$, Fig.~\ref{4}f) towards the reconstructed surface. Since this non-Hermitian effect occurs only on the reconstructed surface (surface 2), Hermitian properties are retained in both the bulk and the unreconstructed surface (surface 1). Consequently, the surface-selective reconstructed Weyl semimetal NdAlSi can be viewed as a Hermitian bulk -- non-Hermitian boundary system, with the NTSFAs considered as non-Hermitian boundary states. The Hermitian bulk -- non-Hermitian boundary system establish in surface-reconstructed Weyl semimetals is markedly different from that in topological insulators\cite{FSchindler_PRXQ2023_KKawabata}, requiring further theoretical research for accurate classification. We refer to the accumulation of non-topological dissipated electrons on the reconstructed surface, due to the non-Hermitian nature of its boundary, as the non-Hermitian boundary electron skin effect (Fig.~\ref{4}f). Under an applied magnetic field, the greater the spectral weight of the NTSFAs and the shorter the quasiparticle lifetime of SFA, the more pronounced the non-Hermitian boundary electron skin effect becomes. 

The non-Hermitian boundary effect is highly sensitive to the boundary of materials and leads to the loss of topological electrons in the Weyl semimetal, impacting its overall topological transport properties. This aligns with the unusual transport measurements observed in NdAlSi especially in the low magnetic field regime\cite{JGaudet_NM2021_CLBroholm,JFWang_PRB2022_GFChen,HYYang_PRM2023_FTafti}. In NdAlSi, the broadening of the SFA due to non-Hermitian effects on the reconstructed surface can weaken their connection to the Weyl points, increasing scattering, particularly in the low-field regime where carriers are more easily deflected. Moreover, some topological electrons are scattered into the NTSFA, transforming into non-topological electrons, which further diminishes the topological transport properties. As a result, the reduced coherence of quasiparticles associated with the Fermi arcs and the loss of topological electrons in the loop, caused by non-Hermitian effects from surface reconstruction, lead to quantum oscillation damping in the low-field regime and unusual behaviors of magnetoresistance\cite{JGaudet_NM2021_CLBroholm,JFWang_PRB2022_GFChen} and an anomalous Hall effect that deviates from typical magnetization curves and may even vanish\cite{HYYang_PRM2023_FTafti}. Interestingly, this unusual transport behavior is absent in NdAlGe\cite{KCho_MC2023_JSRhyee,HYYang_PRM2023_FTafti,CDhital_PRB2023_JFDiTusa,NKikugawa_PRB2024_TTerashima}, despite NdAlSi and NdAlGe belonging to the same material family, sharing similar helical magnetic structures\cite{JGaudet_NM2021_CLBroholm,HYYang_PRM2023_FTafti,CDhital_PRB2023_JFDiTusa} and bulk band structures\cite{JGaudet_NM2021_CLBroholm,CLi_NC2023_OTjernberg,CLi_arxiv2024_OTjernberg,HYYang_PRM2023_FTafti}. The key distinction lies in the cleavage planes: NdAlSi cleaves along the Nd-Al plane\cite{CLi_NC2023_OTjernberg,CLi_arxiv2024_OTjernberg}, while NdAlGe cleaves along the Nd-Ge plane. This is because the Nd-Al and Nd-Ge bonds are the longest in their respective lattices, making them easier to break. In NdAlSi, surface reconstruction occurs on the Al-terminated surface due to the covalent nature of its bonds. In contrast, both the Nd- and Ge-terminated surfaces of NdAlGe exhibit metallic bonding characteristics, making surface reconstruction less likely to occur. Thus, we can reasonably infer that the non-Hermitian effects caused by surface reconstruction of the Al-terminated surface in NdAlSi are responsible for the unusual transport behaviors observed\cite{JGaudet_NM2021_CLBroholm,JFWang_PRB2022_GFChen}, as reported in contrast to NdAlGe\cite{KCho_MC2023_JSRhyee,HYYang_PRM2023_FTafti,CDhital_PRB2023_JFDiTusa,NKikugawa_PRB2024_TTerashima}. This indicates that merely changing the surface structure or surface electronic structures can affect the transport properties of the entire Weyl semimetal. Therefore, it is possible to effectively switch the transport properties by preparing Weyl semimetal NdAlSi films with or without surface reconstruction, making Weyl semimetal NdAlSi a promising candidate for surface sensor applications. 


Thus, we have demonstrated the existence of surface-selective spontaneous reconstruction in the noncentrosymmetric magnetic Weyl semimetal NdAlSi. We demonstrate that surface reconstruction in NdAlSi introduces non-Hermitian terms into the effective Hamiltonian for surface states, which alters the connectivity of SFA and creates new isolated SFAs. Unlike the original SFAs, these new NTSFAs do not connect Weyl nodes of opposite chirality and further hybridize with bulk states. The non-Hermitian boundary effects on reconstructed surface causes the SFA to broaden compared to the unreconstructed surface. In the presence of a magnetic field, the NTSFA acts as a loss mechanism and OBC for the topological electrons and bulk states, providing a good platform for exploring new physical phenomena and potential applications based on boundary non-Hermitian effects. The discovery opens the path to an experimentally viable framework for investigating non-Hermitian effects in condensed matter physics, transcending purely mathematical concepts. As such, it can potentially also pave the way for novel concepts and applications in electronics and photonics.\\


\noindent {\bf Methods}\\
\noindent{\bf Sample} Single crystals of NdAlSi were grown from Al as flux. Nd, Al, Si elements were sealed in an alumina crucible with the molar ratio of 1: 10: 1. The crucible was finally sealed in a highly evacuated quartz tube. The tube was heated up to 1273 K, maintained for 12 hours and then cooled down to 973 K at a rate of 3 K per hour. Single crystals were separated from the flux by centrifuging. The Al flux attached to the single crystals were removed by a dilute NaOH solution.

\noindent{\bf ARPES Measurements} High-resolution ARPES measurements were performed at the Bloch beamline of MAX IV and at the I05 beamline of the Diamond synchrotron light source. The total energy resolution (analyzer and beamline) was set at 15$\sim$20 meV for the measurements. The angular resolution of the analyser was $\sim$0.1 degree. The beamline spot size on the sample was about 10 $\mu$m$\times$12 $\mu$m at the Bloch beamline of MAX IV and about 70 $\mu$m$\times$70 $\mu$m at the I05 beamline of the Diamond synchrotron. The samples were cleaved {\it in situ} and measured at about 18 K at the Bloch beamline of MAX IV and about 10 K at the I05 beamline of the Diamond synchrotron in ultrahigh vacuum with a base pressure better than 1.0$\times$10$^{-10}$ mbar.

\noindent{\bf STM Measurements} STM experiments were conducted using an ultrahigh vacuum (UHV) low-temperature STM (Unisoku, USM1300) under a base pressure below 1$\times$10$^{-10}$ mbar. High-quality NdAlSi single crystals, measuring up to 3 mm$\times$3 mm$\times$3 mm, were bisected prior to being affixed to beryllium copper sheets. The upward-facing surfaces of these halves, which are opposite along the c-axis, provide different terminal surfaces. These crystals were mechanically cleaved in situ at 78 K and promptly placed into the microscope head, which was already at the base temperature of He4 (4.7 K). For high-temperature experiments, the microscope head was heated to 10 K via a Lake Shore Model350 temperature controller, achieving an accuracy of 0.1 K. In contrast, for ultra-low temperature experiments, the scanning head was cooled to 0.3 K using a He3-based single-shot refrigerator, which can be maintained for 100 hours. Topographic images were obtained using Ir/Pt tips in constant-current mode.

\noindent{\bf DFT calculations} The surface states of NdAlSi were studied by constructing a 36 atomic layers slab system with a 20 \AA\ vacuum layer. The electronic structure calculations for this slab system were performed based on the DFT\cite{PHohenberg_PR1964_WKohn,WKohn_PR1965_LJSham} as implemented in the VASP package\cite{GKresse_CMS1996_JFurthmuller,GKresse_PRB1996_JFurthmuller}. The generalized gradient approximation (GGA) of Perdew-Burke-Ernzerhof (PBE) type\cite{JPPerdew_PRL1996_MErnzerhof} was chosen for the exchange-correlation functional. The projector augmented wave (PAW) method\cite{PEBlochl_PRB1994,GKresse_PRB1998_DJoubert} was adopted to describe the interactions between valence electrons and nuclei. The spin-orbit coupling effect was also included. The kinetic energy cutoff of the plane-wave basis was set to be 350 eV. A 12$\times$12$\times$1 Monkhorst-Pack grid\cite{HJMonkhorst_PRB1976_JDPack} was used for the BZ sampling. For describing the Fermi-Dirac distribution function, a Gaussian smearing of 0.05 eV was used. The surface atomic reconstruction was confirmed by the phonon instability of a 18 atomic layers system with a 20 \AA\ vacuum layer, as shown in the Fig. 3l. The phonon spectra were studied within the framework of density functional perturbation theory\cite{SBaroni_RMP2001_PGiannozzi} as implemented in the Quantum ESPRESSO (QE) package\cite{PGiannozzi_JPCM2009_RMWentzcovitch}. In the phonon calculation, a 4$\times$4$\times$1 $\bf q$-mesh was adopted. As a result, in Fig. 3l, we found the phonon mode with the greatest negative frequency around the $\overline{Y}$($\overline{X}$) point, whose vibration mode corresponds to the surface Al atomic distortion in a 2$\times$1 supercell. This is consistent with our STM observations in Fig. 3m.

\noindent {\bf Data Availability}

\noindent The authors declare that all data supporting the findings of this study are available within the paper and its Supplementary Information files.

\vspace{3mm}

\noindent {\bf Acknowledgement}\\
We thank Zhesen Yang for his useful discussion of non-Hermitian physics. The work presented here was financially supported by the Swedish Research council (2019-00701) and the Knut and Alice Wallenberg foundation (2018.0104). Y.G.S. acknowledges the National Natural Science Foundation of China (Grants No. U2032204), and the Informatization Plan of Chinese Academy of Sciences (CAS-WX2021SF-0102). We acknowledge MAX IV Laboratory for time on Beamline BLOCH under Proposal 20221199, 20230262 and 20231119. Research conducted at MAX IV, a Swedish national user facility, is supported by the Swedish Research council under contract 2018-07152, the Swedish Governmental Agency for Innovation Systems under contract 2018-04969, and Formas under contract 2019-02496.

\vspace{3mm}

\noindent {\bf Author Contributions}\\
C.L. proposed and conceived the project. C.L. carried out the ARPES experiments with the assistance from Y.W. and W.Y.C.. J.F.Z. and T.X. contributed to the band structure calculations and theoretical discussion. H.X.L. and Y.G.S. contributed to NdAlSi crystal growth. G.W.L., H.B.D. and J.X.Y. carried out the STM experiments. C.L. contributed to software development for data analysis and analyzed the data with Y.W.. C.L. wrote the paper. C.P., B.T. and T.K. provided the beamline support. Y.W. and O.T. participate in the scientific discussions. O.T. revised the manuscript. All authors participated in and commented on the paper.

\noindent {\bf Competing Interests}\\
The authors declare no competing interests.

\newpage

\end{document}